\documentclass[preprint]{aastex}
\usepackage{txfonts}

\usepackage{tabularx}
\usepackage{graphicx}

\usepackage{natbib}

\newcommand{\mic}{\,{\rm \mu m} } 
\newcommand{\YBG}{\rm Y_{BG}}
\newcommand{\YVSG}{\rm Y_{VSG}}
\newcommand{\YPAH}{\rm Y_{PAH}}
\newcommand{\XISRF}{\rm X_{ISRF}}
\newcommand{\DOR}{\rm 30-Doradus\,}

\begin{document}



\title{Spatial variations of dust abundances across the
Large Magellanic Cloud}
\author{D\'eborah Paradis \altaffilmark{1} \altaffiltext{1}{Infrared Processing and Analysis Center, California Institute of Technology, Pasadena, CA 91125, USA}, William T. Reach \altaffilmark{1}, Jean-Philippe Bernard \altaffilmark{2,3}\altaffiltext{2}{Universit\'e de Toulouse; UPS; CESR; 9 av. du Colonel Roche, F-31028 Toulouse cedex 9, France}\altaffiltext{3}{CNRS; UMR5187; F-31028 Toulouse, France},
Miwa Block \altaffilmark{4} \altaffiltext{4}{Steward Observatory, University of Arizona, 933 North Cherry Ave., Tucson, AZ 85721, USA},
Chad W. Engelbracht \altaffilmark{4},
Karl Gordon \altaffilmark{4},
Joseph L. Hora \altaffilmark{5} \altaffiltext{5}{Harvard-Smithsonian Center for Astrophysics, 60 Garden St., MS-65, Cambridge, MA, 02138-1516, USA},
Remy Indebetouw \altaffilmark{6} \altaffiltext{6}{7 National Radio Astronomy Observatory and Department of Astronomy, University of Virginia, P.O. Box 3818, Charlottesville, VA 22903-0818, USA},
Akiko Kawamura \altaffilmark{7} \altaffiltext{7}{Nagoya University, Dept. of Astrophysics, Chikusa-Ku, Nagoya, 464-01, Japan},
Marilyn Meade \altaffilmark{8} \altaffiltext{8}{Department of Astronomy, University of Wisconsin, Madison, 475 N. Charter St., Madison, WI 53706, USA},
Margaret Meixner \altaffilmark{9} \altaffiltext{9}{Space Telescope Science Institute, 3700 San Martin Drive, Baltimore, MD 21218, USA},
Marta Sewilo \altaffilmark{9},
Uma P. Vijh \altaffilmark{10} \altaffiltext{10}{Ritter Astrophysical Research Center, University of Toledo, Toledo, OH 43606, USA},
Kevin Volk \altaffilmark{11} \altaffiltext{11}{Gemini Observatory, Northern Operations Center, 670 N. Aohuku Place, Hilo, HI 96720, USA}}

\begin{abstract}

Using the data obtained with the Spitzer Space telescope as part of
the Surveying the Agents of a Galaxy's Evolution (SAGE) legacy survey,
we have studied the variations of the dust composition and abundance
across the Large Magellanic Cloud (LMC). Such variations are expected,
as the explosive events which have lead to the formation of the many
HI shells observed should have affected the dust properties. Using a
model and comparing with a reference spectral energy distribution from our Galaxy, we deduce the relative abundance variations of
small dust grains across the LMC. We examined the infrared color
ratios as well as the relative abundances of very small grains (VSGs)
and polycyclic aromatic hydrocarbons (PAHs) relative to the big grain
(BG) abundance. Results show that each dust component could have
different origins or evolution in the interstellar medium (ISM).  The
VSG abundance traces the star formation activity and could result from
shattering of larger grains, whereas the PAH abundance increases
around molecular clouds as well as in the stellar bar, where they
could have been injected into the ISM during mass loss from old stars.

\end{abstract}

\keywords{Magellanic Clouds -- infrared: ISM -- dust, extinction -- ISM: abundances}

\section{Introduction}

The Large Magellanic Cloud (LMC) is the nearest galaxy external to the
Milky Way (MW), located at a distance of about 55 kpc, and is oriented with
an advantageous almost face-on viewing angle
\citep[35$\degr$,][]{vandermarel01}. The LMC is a dwarf galaxy orbiting
around the MW, containing more than 30 billion stars. As
opposed to the situation in the Galaxy, where studies of the infrared
emission from dust in the interstellar medium (ISM) suffers from
significant confusion along the line of sight, the favorable viewing
angle for the LMC offers a direct view to the processes taking place
in the diffuse ISM. The LMC has a lower metallicity than that of our
Galaxy, with estimated values close to 0.3-0.5 Z$\rm _{\odot}$
\citep{Westerlund97} or 0.25 Z$\rm _{\odot}$ \citep{Dufour84} ;
specifically for carbon and oxygen, Z$\rm _{LMC}^C$ = 0.28
Z$\rm_{\odot}^C$ and Z$\rm _{LMC}^O$ = 0.54 Z$\rm_{\odot}^O$.

Extinction curves in several regions of the LMC
\citep{Clayton85,Sauvage91,Gordon03} also differ than those measured
in the Galaxy. For example, the ultraviolet (UV) bump at 2175 $\rm \AA$ is weaker in
the LMC than in the MW. Moreover, the far-ultraviolet (FUV) rise is steeper in the
LMC, indicating that grains responsible for this feature could be more
abundant. However, the LMC and MW extinction curves are similar
in the visible (VIS) and the near-infrared (NIR). The $\rm N_{HI}/Av$ ratio values
\citep{Gordon03,Cox06} indicate a gas/dust ratio ranging between two and
four times the Galactic value, depending on the line of
sight. \cite{Sakon06} analyzed the LMC far-infrared (FIR) emission
using DIRBE data and showed the existence of some pixels with very
bright emission at 240 $\mic$ which could possibly be interpreted as
cold dust. Emission in DIRBE bands at 12 and 25 $\mic$ normalized to
FIR brightness was found to be relatively lower in the LMC
than in our Galaxy. These results can be interpreted as a polycyclic
aromatic hydrocarbon (PAH) and very small grain (VSG) deficit in the
LMC. \cite{Sauvage91} studied the Infrared Astronomical Satellite (IRAS) I$_{12}$/I$_{25}$ versus
I$_{60}$/I$_{100}$ ratio in the LMC and SMC and compared them with
observations of several galaxies by \cite{Helou86}. They
showed that there is a decrease of the $12 \mic$ emission relative to
the 60 and 100 $\mic$ emission associated with the decrease of the
metallicity in the LMC and SMC compared to the MW. They
proposed the low metallicity in the LMC and SMC could limit small
grain formation, concluding that the metallicity could act
preferentially on the abundance of one grain component.

\cite{Bernard08} performed a global preliminary analysis of the
entire LMC using new Spitzer images \citep{Meixner06} covering the
entire LMC with much higher angular resolution than the previous IRAS
or DIRBE data.  They produced a temperature map for the overall LMC
and found temperature variations in the range 12.1 - 34.7 K. They
studied three individual regions, determining and modeling their
spectral energy distribution (SEDs). They showed that the relative abundance of the dust components
is close to that in our Galaxy, in contrast to the previous IRAS and
DIRBE studies.  They also found that the PAH and VSG abundances are
higher in the molecular phase than in the neutral phase for a few
selected regions.  They explained this result as an increase of the
two types of grains in unresolved halos surrounding molecular
clouds. They also evidenced the presence of an excess of the mid-infrared (MIR)
emission in the LMC (70 $\mic$ excess) which is localized mostly in the
atomic medium and in a large circular region around \DOR. They modeled
the excess using a modified size distribution for the VSGs and
interpreted it as a possible sign of shattering of larger dust grains
into smaller VSGs.

Even if we can interpret infrared color variations as abundance
variations, we do not know in general what the origin of these
variations is, nor where the different types of grains are
formed. Grains could be destroyed in supernova shocks, but grain
growth probably takes place in dense molecular clouds through grain
coagulation. It is widely accepted \citep[e.g.,][]{Allamandola89,
Matsuura08}, that PAH molecules could be formed in the expanding
envelopes of carbon-rich stars. The objective of our work is to trace
dust abundance variations across the LMC in order to gather
information about the dust formation and evolution in the ISM.

Sections \ref{sec:obs} and \ref{sec:treatment} briefly summarize the
observations and the processing we applied to the Spitzer images. In
Section \ref{sec:colors} we use an empirical method to study the infrared
color ratios. In Section \ref{sec:ab}, we focus on the dust abundance
variations across the LMC using a dust emission model. Sections
\ref{sec:discussion} and \ref{sec:cl} are devoted to discussion and
conclusion.

\section{LMC observations}
\label{sec:obs}
\subsection{Spitzer Space Telescope Observations}

The Spitzer Space Telescope \citep{Werner04} carries three
instruments: the Infrared Array Camera \citep[IRAC,][]{Fazio04} which
provides imaging in the near and MIR domain (3.6, 4.5, 5.8
and 8.0 $\mic$), the Infrared Spectrograph \citep[IRS,][]{Houck04}
which provides both high- and low-resolution spectroscopy at
MIR wavelengths, and the Multiband Imaging Photometer (MIPS) for
Spitzer \citep[][]{Rieke04} providing imaging and limited
spectroscopic capabilities at three FIR wavelengths (24, 70
and 160 $\mic$). IRAC data have an angular resolution of 1$_{.}^{\prime
\prime}$6 - 1$_{.}^{\prime \prime}$9 from 3.6 to 8 $\mic$, and MIPS data
have an angular resolution of 6$^{\prime \prime}$, 18$^{\prime
\prime}$, and 40$^{\prime \prime}$ at 24, 70, and 160 $\mic$
respectively.

The data were obtained as part of the Surveying the Agents of
a Galaxy's Evolution (SAGE) Spitzer legacy survey \citep{Meixner06}.  The
survey covered a $\sim 7\times7$ deg region centered on the LMC
with all of the IRAC and MIPS bands.  We used mosaics which combined
both epochs of the data into a final map at a resolution of
3$_{.}^{\prime \prime}$6 pixel$^{-1}$ for IRAC, and 2.49, 4.8, and 15$_{.}^{\prime
\prime}$6 pixel$^{-1}$ for the MIPS 24, 70, and 160 $\mic$ maps, respectively.

\subsection{IRIS data}

The IRIS data are the improved version of latest IRAS images, corresponding to the second generation
processing (or IRAS Sky Survey Atlas, ISSA), see \cite{MamD05}.
These maps have been reprocessed to improve sensitivity and absolute
calibration. Compared to the latest version of the IRAS data, this new
product named IRIS (for Improved Reprocessing of the IRAS Survey)
benefits from a better zodiacal light subtraction, calibration and
zero level adjustments to match the DIRBE data on large angular
scales. With a resolution near 4$^{\prime}$ the IRIS survey covers $98
\%$ of the sky at four wavelengths (12, 25, 60 and 100 $\mic$).

\section{Image Processing}
\label{sec:treatment}

\begin{figure*}
\begin{center}
\includegraphics[width=16.5cm]{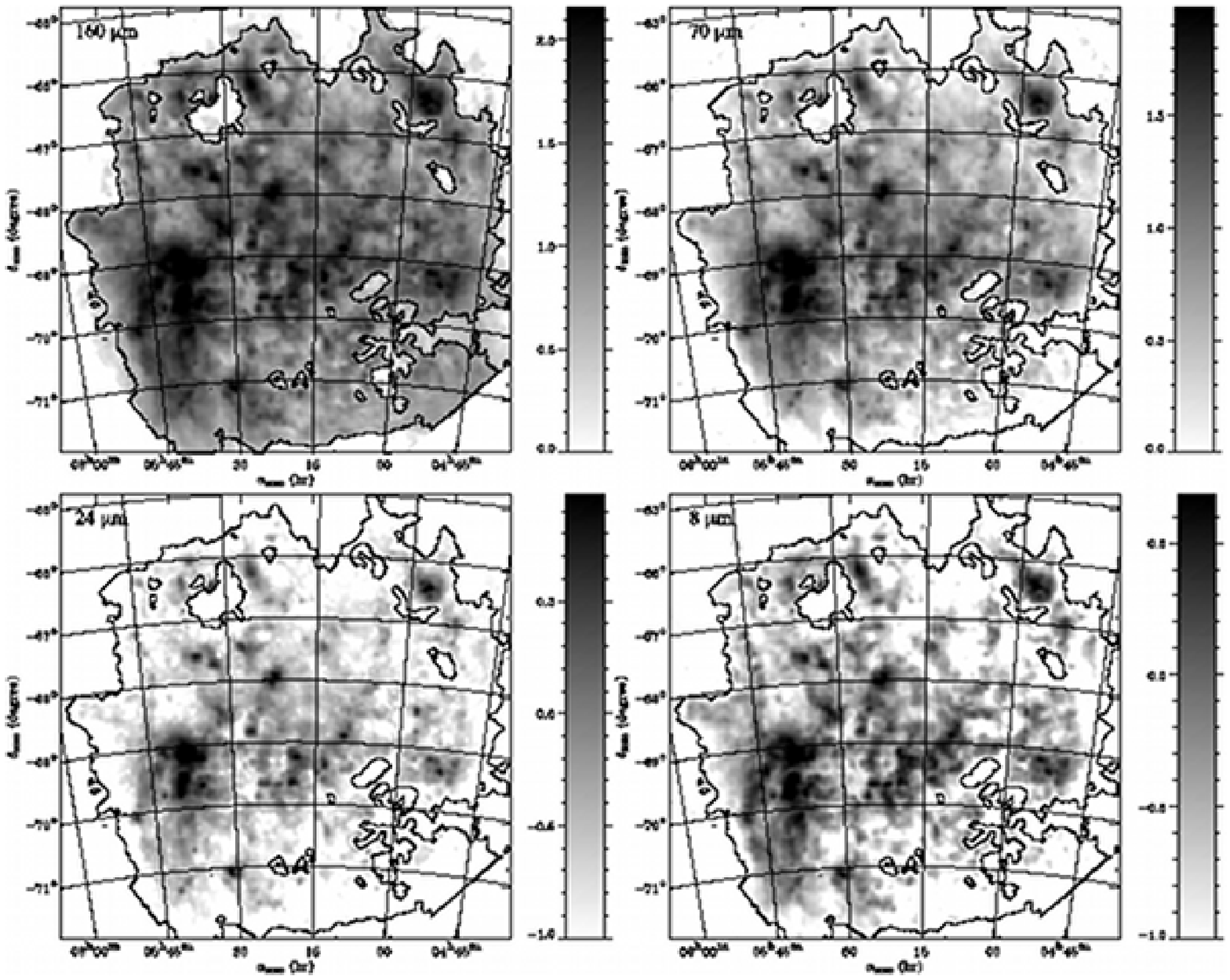}
\caption{Spitzer images at 160, 70, 24, and 8 $\mic$, at the
4$^{\prime}$ angular resolution, after the processing described in
Section \ref{sec:treatment}. The grayscale is log$\rm
_{10}$(I$\rm _{\nu}$(MJy sr$^{-1}$)).\label{fig_spitzer} }
\end{center}
\end{figure*}

\begin{figure*}
\begin{center}
\includegraphics[width=16.5cm]{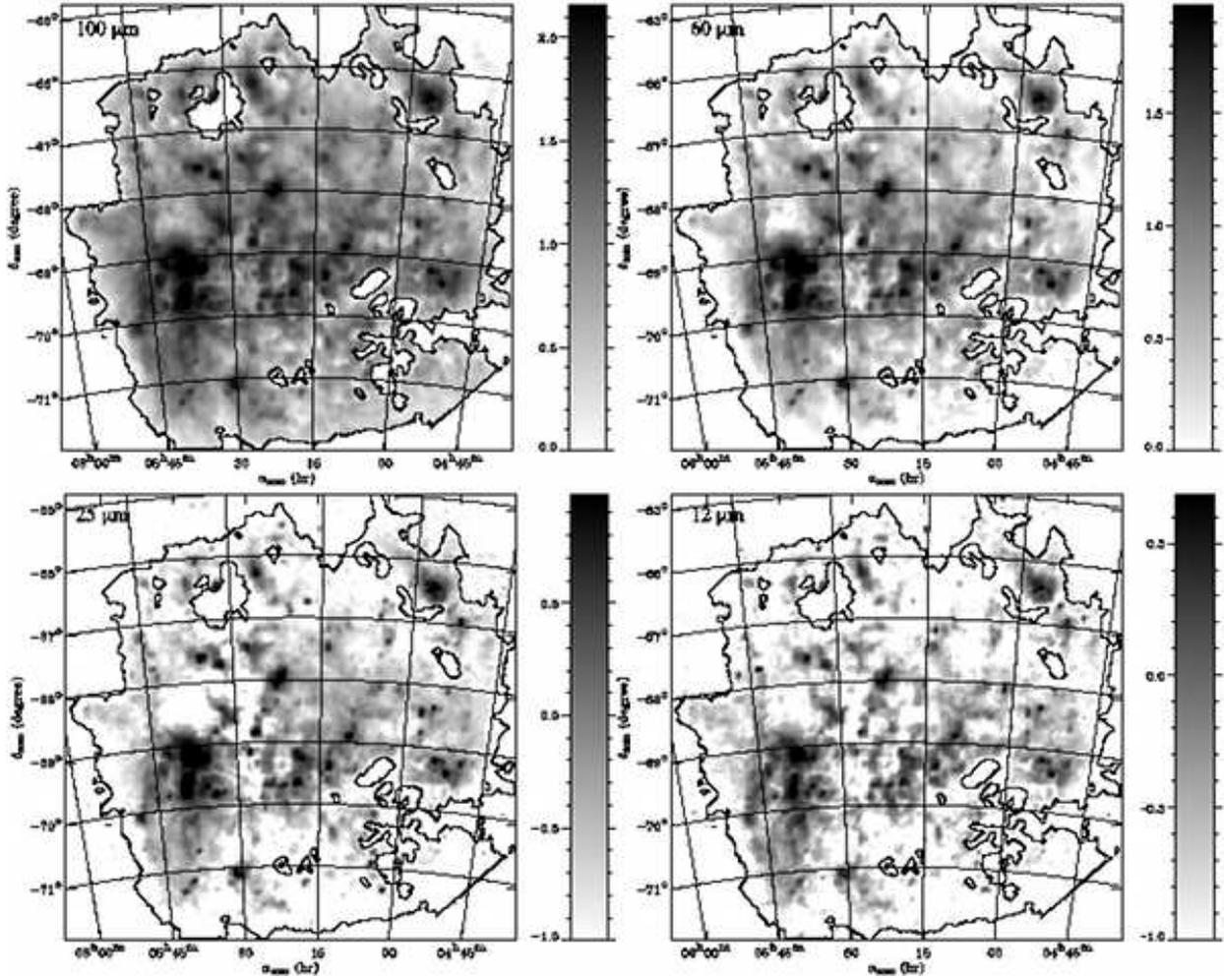}
\caption{IRIS images at 100, 60, 25, and 12 $\mic$, at the 4$^{\prime}$
angular resolution, after the processing described in Section
\ref{sec:treatment}. The grayscale is log$\rm
_{10}$(I$\rm _{\nu}$(MJy sr$^{-1}$)).\label{fig_iras} }
\end{center}
\end{figure*}

We briefly describe here the processing we applied to each
map. Details of the data treatment used here are given in
\cite{Bernard08}. All data were projected on a common grid centered
at $\rm \alpha_{2000}=05^h18^m48^s$ and $\rm
\delta_{2000}=-68\degr42^{\prime}00^{\prime \prime}$, with a pixel
size of $\rm 2^{\prime}$ and then smoothed to the IRIS resolution of
$\rm 4^{\prime}$. We subtracted the Galactic foreground emission in
order to avoid any confusion between the emission from the LMC and
from our Galaxy. We used the SED of high
latitude galactic emission of \cite{Dwek97} \citep[see][]{Bernard08},
and the HI 21 cm emission in the heliocentric velocity range $\rm
-64<v_{hel}< 100$ km s$^{-1}$ \citep{Staveleysmith03}. In order to
eliminate residual zodiacal emission and cosmic infrared background
(CIB), we have subtracted a background from all images corresponding
to the median intensity in the region located 4$\rm \degr$ from the
center of the maps. Point sources in archival catalogs (Two Micron All Sky Survey (2MASS) catalog for IRAC and SAGE catalog for MIPS) were masked from the IRAC and MIPS
maps before degrading their resolution to $\rm 4^{\prime}$, to match that of
the IRIS maps. For the IRIS point-source subtraction we used the SAGE
catalogs; specifically the 160, 70, 24 and 8 $\mic$ catalogs for the
IRIS 100, 60, 25 and 12 $\mic$ maps respectively.  The Spitzer and
IRIS images at the 4$^{\prime}$ angular resolution are presented in
Figures \ref{fig_spitzer} and \ref{fig_iras}.

\section{Infrared colors}
\label{sec:colors}

\subsection{Color-Color diagrams}
\label{sec:cc_diag}

\begin{figure*}
\begin{center}
\includegraphics[width=16.5cm]{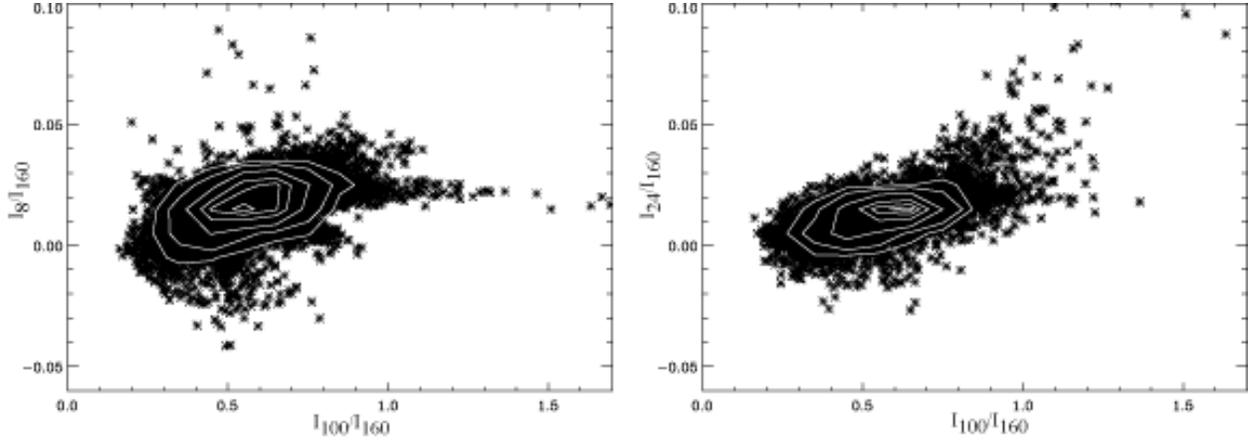}
\caption{I$_8$/I$_{160}$ $\mic$ (left) and I$_{24}$/I$_{160}$
$\mic$ (right) vs. I$_{100}$/I$_{160}$ $\mic$ ratios. Every
group of three points has been averaged to produce one plotted
point. \label{fig_cc} }
\end{center}
\end{figure*}

We first examine the trends in the LMC color-color diagrams (see
Figure \ref{fig_cc}). The I$_8$/I$_{160}$ and I$_{24}$/I$_{160}$
ratios can be considered as proxies for the PAH and VSG relative abundances
with respect to that of the big grains (BGs) respectively. The
I$_{100}$/I$_{160}$ ratio is directly linked to BG equilibrium temperature and
therefore to the interstellar radiation field (ISRF) intensity. The
I$_8$/I$_{160}$ and I$_{24}$/I$_{160}$ ratios show a trend with the
I$_{100}$/I$_{160}$ ratio, such that the ratios increase with the BG
temperature. However, in the first case the dispersion is quite large,
whereas in the second case the trend is more pronounced. We can
conclude that the PAH abundance does not depend strongly on the radiation
field intensity.
The VSG abundance, however, seems to be increased in warm regions
of the LMC.

\begin{figure*}
\begin{center}
\includegraphics[width=16.5cm]{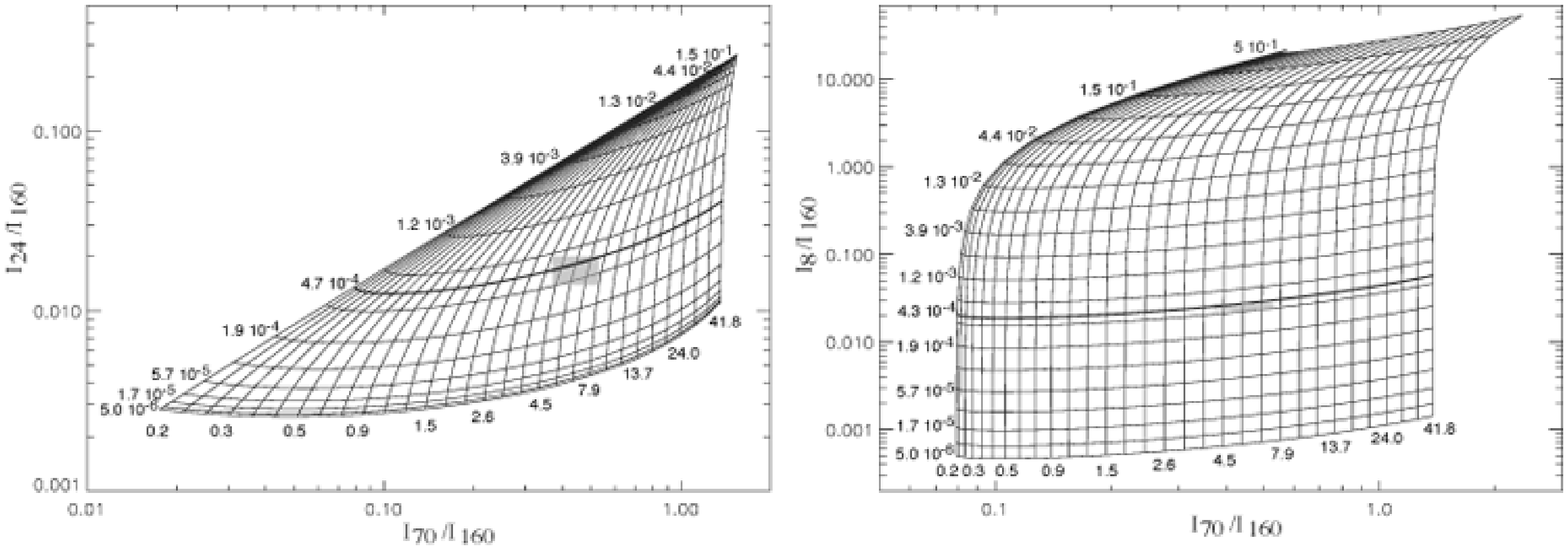}
\caption{Predicted I$_{24}$/I$_{160}$ (left) and
I$_{8}$/I$_{160}$ (right) vs. I$_{70}$/I$_{160}$ using
the \cite{Desert90} model, for various intensities of the ISRF (from
0.2 to 42) and VSG mass abundances (left) and PAH mass abundances
(right) (from $\rm Y_{VSG,PAH}=5\times10^{-6}$ to $\rm
Y_{VSG,PAH}=0.5$). The bold lines correspond to the reference MW
dust abundances from \cite{Desert90} : $\rm Y_{VSG}=4.7 \times 10^{-4}$ and
$\rm Y_{PAH}=4.3 \times 10^{-4}$. The gray rectangular boxes show the color values derived
for the LMC (see the text). The size of the boxes shows the error. \label{fig_dustem} }
\end{center}
\end{figure*}

To connect the observed brightnesses to dust abundances more
precisely, we use an updated version of the \cite{Desert90} model,
which includes an improved description of the PAH IR emission bands at
3.3, 6.2, 7.7, 8.6, 11.3, 12.7 and 16.4 $\mic$. This version also
takes into account the results of \cite{Flagey06}, based on the work
by \cite{Rapacioli05} including the cross sections of neutral and
ionized PAHs. We assume the same ISRF spectral shape as in the solar
neighborhood.
We show in Figure \ref{fig_dustem} the variations of the predicted
I$_{24}$/I$_{160}$ and I$_{8}$/I$_{160}$ with I$_{70}$/I$_{160}$ for
different values of ISRF intensity ($\XISRF$), from 0.2 to about 42, and various VSG
and PAH mass abundances relative to hydrogen. The reference MW
dust abundances from \cite{Desert90} are $\YBG=6.4\times10^{-3}$,
$\YVSG=4.7\times10^{-4}$, $\YPAH=4.3\times10^{-4}$, where $\rm Y_X$ is the
mass of dust species X relative to H. In Figure \ref{fig_dustem}, we
allowed the $\YVSG$ and $\YPAH$ parameters to vary from 5 $\times10
^{-6}$ to 0.5.  Comparing the values of the I$_{8}$/I$_{160}$, I$_{24}$/I$_{160}$ and
I$_{70}$/I$_{160}$ ratios given in Table\,\ref{tab_ir_color} to the
predicted values deduced from the model, we deduce from
Figure\,\ref{fig_dustem} that the intensity of the ISRF in the diffuse
LMC medium is about  $\rm 4.5<X_{ISRF}<6.5$, taking into account the
errors on the brightness ratios.  Similarly, the values of the VSG and
PAH mass abundance are in the range $\rm 3.5\,10^{-4}<\YVSG<4.7\,10 ^{-4}$
and $\rm 3.5\,10^{-4}<\YPAH<4.3\,10 ^{-4}$ respectively.  The diffuse ISM
in the LMC is therefore subjected to an ISRF which is moderately
stronger than that in the solar neighborhood, and appears to contain a
similar amount of VSGs and PAHs with respect to large grains, once the
effect of heating has been properly taken into account.

\subsection{Infrared correlations}

\begin{figure*}
\begin{center}
\includegraphics[width=11cm]{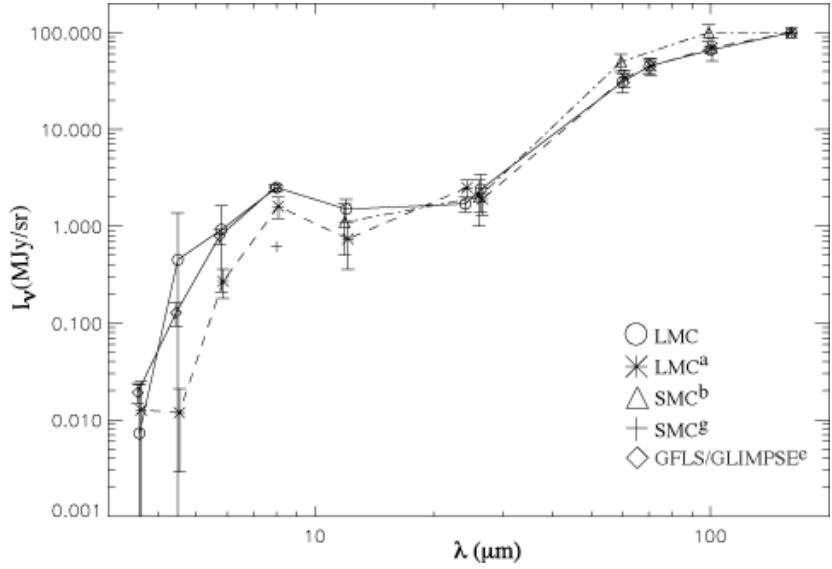}
\caption{SEDs deduced from the IR ratios given in Table
\ref{tab_ir_color}, for the LMC. Our results are represented by
circles. The SED of the whole LMC by \cite{Bernard08} is shown by
stars. The SMC points are from Bot et al. (2004; triangles) and
Bolatto06 et al. (2007; cross). The Galactic diffuse ISM points are from Flagey06 et al. (2006; diamonds). A slight shift in wavelengths has been
applied for clarity.  \label{fig_sed}}
\end{center}
\end{figure*}

In order to measure the color ratios, we performed linear correlations
between the IR brightnesses over the whole LMC, using 
\begin{equation}
I(\lambda_1)=a_1I(\lambda_2)+b_1
\end{equation}
\begin{equation}
I(\lambda_2)=a_2I(\lambda_1)+b_2
\end{equation}
where $\rm a_1$ and $\rm a_2$ are the color ratio parameters to be
determined and $\rm b_1$ and $\rm b_2$ are constants corresponding to
eventual residual offsets in the images.  To avoid any bias in the
results depending on the order of the axis we consider, we take the
mean value between $\rm a_1$ and $\rm 1/a_2$ and we define the
uncertainty on each ratio as the difference between these two
values. In order to identify outliers in the fit, we performed
iterations of the linear regression and removed pixels distant by more
than 3-$\rm \sigma$ from the linear fit, where $\rm \sigma$ is the
standard deviation of the distance between the data and the fit. This
process reached convergence in about 15 iterations. The results of the
regression are given in Table \ref{tab_ir_color}. A graphical
representation of the table is given in Figure \ref{fig_sed}.
We compare the ratios obtained in the LMC with other ratios calculated in the
SMC, in the Galaxy, and in other galaxies. The ratios from 
\cite{Bernard08} have been deduced from the total SED in the overall
LMC (see their Tables 5 and 7). Therefore in this case, individual HII regions such
as the \DOR region are included, which contribute a large fraction of the total
brightness. For that reason our results presented here are different from those
obtained by \cite{Bernard08}.
The correlation method proposed here is more appropriate to study the
diffuse ISM, in the sense that pixels dominated
by the diffuse medium are given more importance than the ones
dominated by individual HII regions, which are disregarded as
outliers. The quoted color ratios in the SMC are taken from
\cite{Bot04} and have been deduced from Gaussian fits of the
histograms of the ratios. We find the I$_{100}$/I$_{160}$ and
I$_{60}$/I$_{160}$ ratios to be 66$\%$ and 62$\%$ lower, respectively,
in the LMC, compared to the I$_{100}$/I$_{170}$ and I$_{60}$/I$_{170}$
ratios in the SMC. These differences must be due to the average
equilibrium temperature of the large grains in the SMC being higher
than in the LMC. However, the $\rm I_{25}/I_{160}$ and $\rm I_{12}/I_{160}$
ratios, which trace mostly the VSG and PAH abundance relative to BGs
and which to first order do not depend on the BG temperature, are
almost the same in the two galaxies.  \cite{Bolatto06} explored the
southwest region of the bar in the SMC and found an overall $\rm
I_{8}/I_{24}$ ratio close to 0.36, lower by a factor of 4 compared to
our results in the LMC. \cite{Engelbracht05} analyzed this ratio as a
function of the metallicity for a sample of 34 galaxies. They show
that the $\rm I_{8}/I_{24}$ ratio decreases with
metallicity. \cite{Dwek09} observed the same trend analyzing the
emission of several nearby galaxies. They interpreted this as
resulting from a lower PAH abundance in low-metallicity
environments. That could explain why this ratio is smaller in the SMC,
which metallicity is smaller than that of the LMC.  According to
\cite{Engelbracht05}, the mean ratio below one-third solar
metallicity is 0.08 $\pm$ 0.04, using a solar metallicity reference
of 12+log(O/H)=8.7 \citep{Prieto01}.
For galaxies with metallicity comparable to that of the LMC (between
1/3 and 1/2 solar), we expect from \cite[][]{Engelbracht05} a $\rm
I_{8}/I_{24}$ ratio close to 0.4. This is significantly lower than the
median value of 1.6 we derived here for the diffuse emission in the
LMC. The PAH relative abundance therefore seems to be larger in the
diffuse ISM of the LMC than in other galaxies with similar low
metallicity. However, \cite{Bernard08} found a similar ratio as
\cite{Engelbracht05}, which could indicate that the ratio
in external galaxies is affected by the contribution from
major HII region similar to \DOR in the LMC.

The I$_{5.8}$/I$_{8}$ ratio has been studied by \cite{Arendt08} in
the Galactic center and by \cite{Flagey06} in the Galactic diffuse
ISM, as observed within the Galactic First
Look Survey (GFLS) and Galactic Legacy Infrared Mid-Plan Survey
Extraordinaire \citep[GLIMPSE,][]{Churchwell05}. In the case of the
Galactic center, \cite{Arendt08} found an uniform ratio for the
diffuse emission, independent of the 8 $\mic$ brightness. Their median
value of I$_{5.8}$/I$_{8}$=0.37 $\pm$ 0.03 is close to our ratio of
0.33 $\pm$ 0.2. The IRAC ratio obtained by \cite{Flagey06} ranges
from 0.26 to 0.37 and is also in agreement with our result. Therefore, the
I$_{5.8}$/I$_8$ ratio seems to be rather similar, around 0.3, in the
LMC and the MW, independent of metallicity.

We find a $\rm I_{4.5}/I_{8}$ ratio in the LMC equal to 0.18 $\pm$
0.37.  \cite{Flagey06} found values between 3.7$\times10^{-2}$ and
6.5$\times10^{-2}$ in the MW, lower than our result by a factor
of 2.8 and 4.9 respectively. The model we use shows that this ratio is
rather independent of the ionization state of the PAHs, increasing by
only about 10\% when going from fully neutral to fully ionized PAHs. It
is therefore likely that ionization is not the reason for the observed
difference.  However the size distribution of PAHs is likely to affect
this ratio, leading to larger ratio values for smaller PAH
sizes. Therefore, the observed difference could indicate the presence
of smaller PAHs in the LMC than in the MW.

For the $\rm I_{3.6}/I_{8}$ ratio we find in the LMC a value of (2.9
$\pm$ 6.3)$\times10^{-3}$, two to three times lower than the ratio derived
by \cite{Flagey06} (after point-source
subtraction). This ratio is sensitive to PAH ionization and
decreases by 90\% when going from fully neutral to fully ionized PAHs.
Therefore, the observed difference with the MW value could indicate
the presence of ionized PAHs in the LMC.
\cite{Engelbracht05} found that in subsolar
metallicity galaxies, galaxies with high values of $\rm I_{4.5}/I_{8}$
($\rm > 0.1$) tend to have low values of $\rm I_{8}/I_{24}$
(essentially below 0.1). In the LMC we have both large $\rm
I_{8}/I_{24}$ and large $\rm I_{4.5}/I_{8}$ ratios and a low
metallicity compared to the solar value. The differences could be due
to distinct properties of the near-infrared continuum or of the
PAHs. Nevertheless, the emission at 3.6 $\mic$ is dominated by stellar
emission, and therefore we caution that point-source subtraction for
these data is difficult and could have an impact on the results.


\section{Dust abundance variations}
\label{sec:ab}
\subsection{Method}
\label{sec:desert}

We consider the IR brightness as the sum of the contributions from
the 3 dust components, BG, VSG, and PAH :
\begin{equation}
\label{eq_lin}
I_{\nu}(\lambda)=Y^{BG}I^{BG}_{\nu}(\lambda)+Y^{VSG}I^{VSG}_{\nu}(\lambda)+Y^{PAH}I^{PAH}_{\nu}(\lambda)
\end{equation}
where $\YBG$, $\YVSG$ and $\YPAH$ correspond to the mass abundances
for each dust component.
The $\rm I_{\nu}$ values for each dust component in equation
\ref{eq_lin} are computed using the version of the \cite{Desert90}
model described in Section \ref{sec:cc_diag}.  The intensities are
computed in the photometric channel of the instruments considered in
the study and have been corrected for the color correction factor,
computed using the filter transmission and flux convention of each
instrument used. The mass abundances are determined by fitting to the
data.

\cite{Bernard08} have shown that a significant emission excess is
present at 70 $\mic$ in the LMC (see their Section 5.3). This excess
has also been found in the SMC by \cite{Bot04}. In the LMC, it is
apparently associated with the neutral phase of the ISM
\citep{Bernard08} and can be attributed to regions of enhanced VSG
abundance. \cite{Bernard08} studied different possible origins to
explain the 70 $\mic$ excess. The most likely explanation
is variations of the VSG size distribution, and the excess
indicates a greater abundance of large VSGs than in our Galaxy. In the
context of the model we use, the VSG size distribution is given by 
\begin{equation}
\frac{dn}{da} \propto a^{-\alpha_{VSG}}
\end{equation}
where dn is the numerical density of grains with a dimension between a
and a+da, and $\rm \alpha_{VSG}$ is the VSG size distribution. In the
Galaxy, \cite{Desert90} derived an average value of
2.6. \cite{Bernard08} found that a value of $\rm \alpha_{VSG}
\thickapprox$1 is required to explain the LMC observations.

\begin{figure*}
\begin{center}
\includegraphics[width=7.5cm]{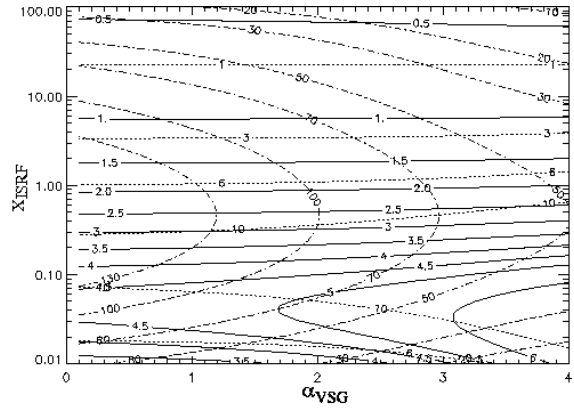}
\caption{I$_{160}$/I$_{100}$ (solid lines), I$_{160}$/I$_{70}$ (dot
lines) and I$_{160}$/I$_{24}$ (dash-dotted lines) deduced from the
\cite{Desert90} model as a function of the intensity of the ISRF ($\rm X_{ISRF}$) and of the index of the VSG size distribution
($\rm \alpha_{VSG}$).\label{fig_contours} }
\end{center}
\end{figure*}

Using the \cite{Desert90} model, we can study the variations of the
different brightness ratios as a function of the $\XISRF$ and $\rm
\alpha_{VSG}$. Figure \ref{fig_contours} shows the evolution of the three
ratios I$_{160}$/I$_{100}$, I$_{160}$/I$_{70}$ and I$_{160}$/I$_{24}$
for different values of the $\XISRF$ and $\rm \alpha_{VSG}$ derived
using the model. The figure shows that $\rm \alpha_{VSG}$ does not
significantly affect the I$_{160}$/I$_{100}$ and I$_{160}$/I$_{70}$
ratios for $\XISRF$ values larger than about 0.1.  For our study, we do
not have enough spectral coverage to warrant extra free parameters, so
we decided to consider only two values of $\rm \alpha_{VSG}$ : the
Galactic value ($\rm \alpha_{VSG}$=2.6) and the value used in Bernard et al. (2008; $\rm \alpha_{VSG}$=1).

As a first step we computed the normalized brightnesses between 160
$\mic$ and 8 $\mic$ (below this wavelength the presence of an
additional NIR continuum is generally needed to explain the
observations) for a grid of $\XISRF$ values and the two selected
values of $\rm \alpha_{VSG}$, using the MW dust abundances from
\cite{Desert90} (see Section \ref{sec:cc_diag}). The brightness for
each dust component i can be written
$\rm I^{i}_{\nu}=I^{i}_{\nu}(\alpha_{VSG}, \XISRF)$. We obtain a table of
brightnesses for the PAHs, VSGs, and BGs separately. We compare the
brightness of our infrared maps with the values in the table at each
pixel of the map, in order to derive maps of the mass abundances. To
ensure that the $\XISRF$ value is calculated correctly, we forced the
model to reproduce the data at 100 and 160 $\mic$ by increasing the
weight of these data points in the SEDs. We derived the mass abundances
minimizing the sum of the squared difference between the model and
observations ($\rm \chi^2$) for each value of $\XISRF$ and $\rm
\alpha_{VSG}$ and selected the solution with minimum $\rm \chi^2$.

One shortcoming of our modeling is the assumption of a single $\rm
X_{ISRF}$ and therefore BG equilibrium temperature, along any line of
sight.  This is expected to be a good approximation in the LMC due to
its near face-on viewing geometry. To account for the possibility of a
mixture of the radiation field intensities along the line of sight, we
follow the description proposed by \cite{Dale01} assuming a power-law
distribution of the dust mass $\rm M_d(X_{ISRF})$ subjected to a given
ISRF intensity on a given line of sight
\begin{equation}
\label{eq_dale}
dM_d(X_{ISRF}) \propto X_{ISRF}^{-\alpha}d\XISRF \,\,\,\,\, 0.3<\XISRF<10^5
\end{equation}
where $\rm \alpha$ is a parameter that represents the relative contribution 
of radiation field intensities, in the range $\rm \alpha$=1-2.5. We first compute 
emission spectra using the \cite{Desert90} model at various $\XISRF$ values 
(noted $\rm  I^{mod}_{\nu}(X_{ISRF})$), then sum them over $\XISRF$ according to 
\begin{equation}
I^{tot}_{\nu}=\frac{\sum_{X_{ISRF}} I^{mod}_{\nu}(X_{ISRF}) \times X_{ISRF}^{-\alpha}}{\sum_{X_{ISRF}} X_{ISRF}^{-\alpha}}
\end{equation}

We follow the same methodology as described previously, constructing a
table of normalized brightnesses for different $\rm \alpha$ values for
each grain component, using the MW abundance values from
\cite{Desert90} as a reference. We use the same values of $\rm
\alpha_{VSG}$ as derived previously and we minimize $\rm \chi^2$ in each
pixel with respect to the above table, determining $\rm \alpha$. The
abundances of the various dust components are then obtained in the
same way as previously.

\subsection{Results}

The objective of our modeling is to determine the distribution of the
VSG and PAH abundance compared to the BG abundance across the LMC. In
order to avoid any confusion between the dust emission and unresolved
stars in the 8 $\mic$ map which could have been left after the initial
8 $\mic$ point-source subtraction, we also masked the point sources
from the initial 3.6 $\mic$ map, and smoothed the map to 4$^{\prime}$. The
median ratio between the point-source fluxes at 3.6 $\mic$ and at 8
$\mic$ is 4.  In order to remove potential contamination by unresolved
stars at $8\mic$, we scaled the 3.6 $\mic$ map by a factor 1/4 and
subtracted it from the 8 $\mic$ map, assuming negligible ISM emission
at 3.6 $\mic$. This is an important step, specially in the stellar bar
where the star density is high and where we identify a higher PAH
abundance derived from the 8 $\mic$ emission map (see
Section \ref{sec_pah}).

Figures \ref{fig_pah_bg} and \ref{fig_vsg_bg} show the spatial
distribution of the PAH and VSG abundances relative to BGs.  It can be
seen that the distributions of these two dust components are quite
different. The white regions around \DOR have very high $\rm \chi^2$
values and have been masked in all the maps. In the following sections,
we analyze separately the variations of the relative abundances of
each dust component.

\subsubsection{PAH distribution}
\label{sec_pah}

\begin{figure*}
\begin{center}
\includegraphics[width=16.5cm]{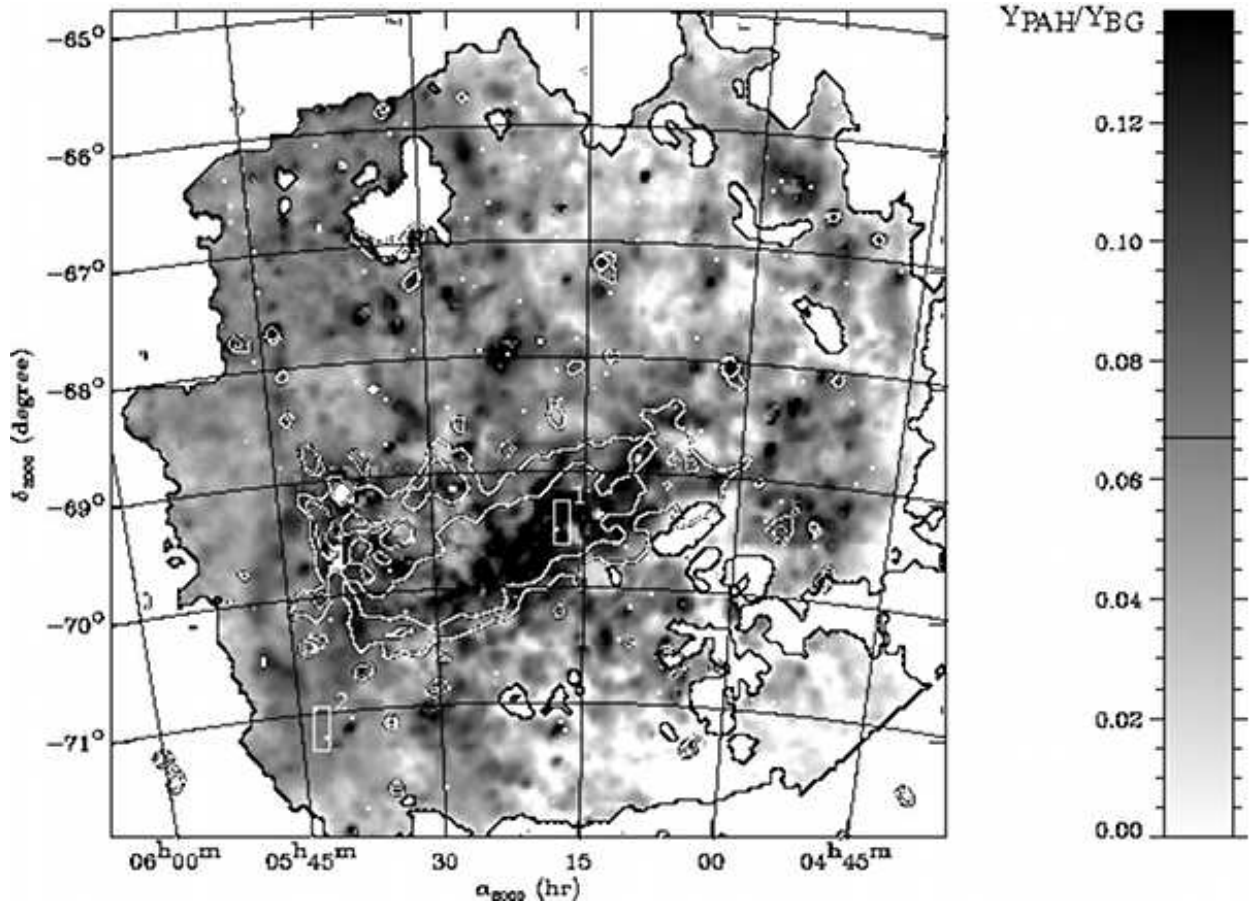}
\caption{Map of the PAH relative abundance with respect to BGs at a
resolution of 4$^{\prime}$, overlaid with contours at 0.5 and 0.7
MJy sr$^{-1}$ of the Two Microns All Sky Survey (2MASS) J-band emission,
convolved to the same resolution. The contours show the extent of the
stellar bar. Regions 1 and 2 indicated by rectangles are discussed in
Section \ref{sec_pah}. The horizontal line in the color bar shows the value of the standard MW relative abundance. \label{fig_pah_bg} }
\end{center}
\end{figure*}

\begin{figure*}
\begin{center}
\includegraphics[width=16.5cm]{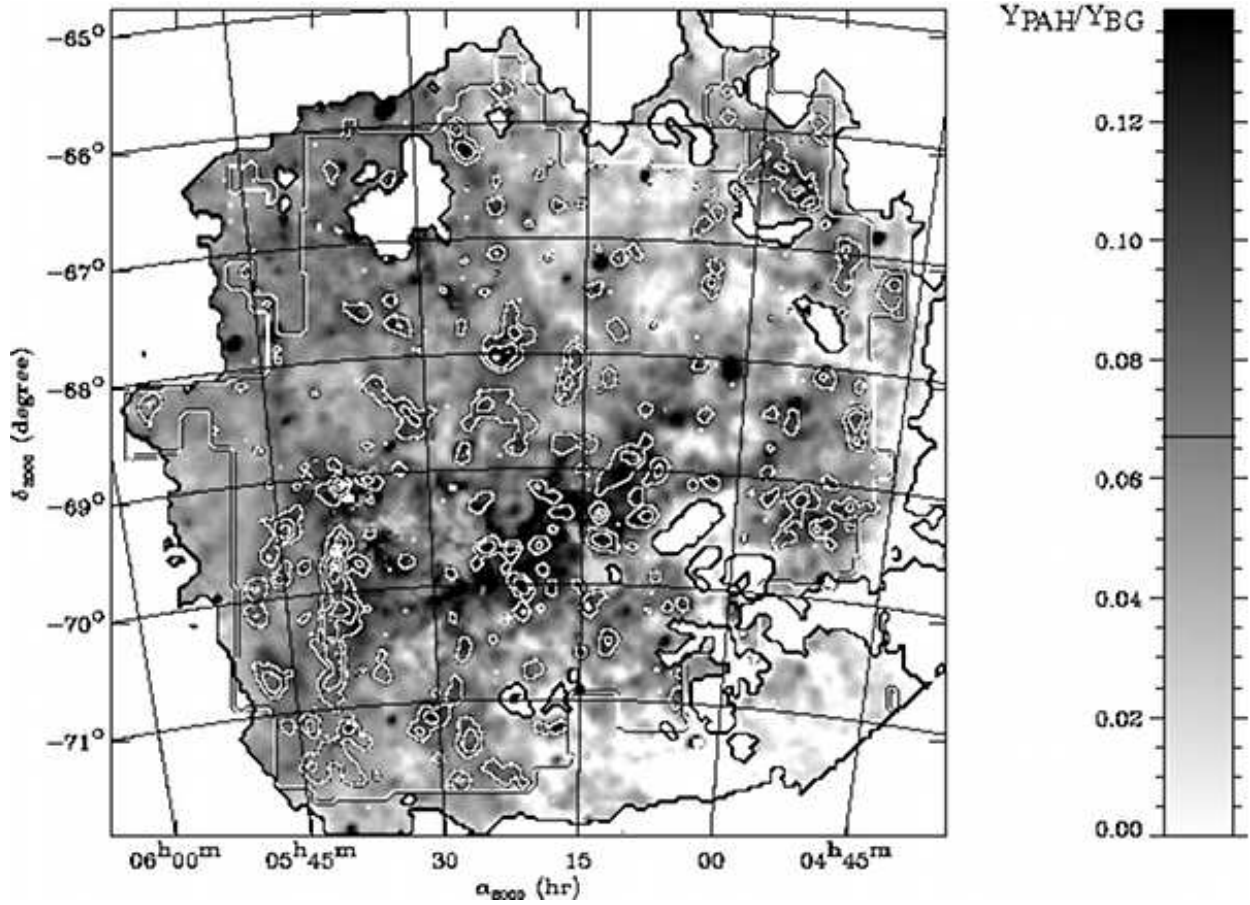}
\caption{Map of the PAH relative abundance with respect to BGs,
overlaid with the NANTEN $^{12}$CO (J=1-0) integrated intensity
contours convolved to the 4$^{\prime}$ resolution, at 0.4, 2 and 6 K km
s$^{-1}$. The lines indicate the limits of the
survey. The horizontal line in the color bar shows the value of the standard MW relative abundance.\label{fig_pah_bg_co} }
\end{center}
\end{figure*}

Figure\,\ref{fig_pah_bg} shows that the PAH relative abundance is
larger than average over a large elongated region toward the center of
the LMC. This region is located within the old-population stellar bar,
delineated in the figure by the 2MASS contours. The PAH relative
abundance increases by a factor of $\sim$2 inside this region,
compared to the rest of the LMC. Figure \ref{fig_pah_bg_co} shows that
the PAH distribution is also sometimes well correlated with the
molecular clouds. This result is in agreement with \cite{Bernard08}
who proposed that this may be due to the presence of enriched PAH
halos similar to those found in the solar neighborhood, mostly
unresolved at the distance of the LMC.  Our results also show
increased PAH abundance toward some HII regions. This could result
from the fact that HII regions are generally associated to molecular
clouds. We also note that HII regions with high $\XISRF$ show low
values of the PAH abundance, in agreement with findings in the MW that
PAHs are deficient in warm HII regions. Note also the presence in
Figure \ref{fig_pah_bg} of a north-south elongated region between \DOR
and the bar (around $\rm \alpha=5^h27^m$) with lower apparent PAH
abundance. This is likely an artifact of the method used. This region
was shown in \cite{Bernard08} to have an increased BG temperature,
which they attributed to an artifact in the IRAS data at 100 $\mic$
following crossing of the bright \DOR region along the IRAS
scans. This results in an overestimated ISRF and an underestimated
PAH abundance. A similar effect is seen on Figure \ref{fig_vsg_bg} for
the VSG abundance.
We also investigated whether the increase of the PAH relative
abundance in the stellar bar could be due to a difference in the ISRF
spectrum compared to what we used in the model. Indeed, the ISRF in
the bar is likely to be dominated by the many old-population stars and
should be redder than in the rest of the LMC. Since BGs are heated
equally by UV and VIS photons, our procedure to derive the radiation
field intensity from their temperature would lead to overestimating
the UV part of the ISRF in that case. Since PAHs are expected to be
mostly sensitive to UV photons, this in turn would lead to
underestimating their abundance. We, however observe an opposite
behavior. We therefore can exclude that the increased PAH abundance
in the stellar bar could be due to this effect.

The top panel of Figure \ref{fig_spec_regions} shows the median SED,
computed in regions 1 and 2 (see Figure \ref{fig_pah_bg}) with low and
high PAH relative abundance ($\rm Y_{PAH}/Y_{BG}=0.076$ and 0.16),
respectively. The SEDs are compared to the prediction of the model
described by Equation \ref{eq_lin}. The error bars correspond to the
standard deviation in the background region (same background region as
described in Section \ref{sec:treatment}) . We took into account the
point-source subtraction in the error estimate by dividing the
standard deviations by a mask whose values ranged between 0 and 1 for
pixels with or without contamination by point sources,
respectively. Table \ref{table_regions} summarizes the coordinates,
size, brightness values, and model predictions for each region. We can
see that the 160 $\mic$ brightness for the two regions are within
15$\%$ of each other, while the difference in brightness at 8 $\mic$
is a factor of 2, which illustrates why the model fits give $\rm
Y_{PAH}/Y_{BG}$ higher by a factor of 2 for region 1. However, we can
see that the model for the two SEDs is not able to reproduce the IRIS
data at 12 $\mic$. The model we consider here assumes a fixed fraction
of neutral and ionized PAHs of 90$\%$ and 10$\%$,
respectively. Therefore we suspect that the difference between the
data and model at 12 $\mic$ could be due to this limitation of the
model. We will discuss this point in Section \ref{sec:discussion}.

\subsubsection{VSG distribution}
\label{sec_vsg}

\begin{figure*}
\begin{center}
\includegraphics[width=16.5cm]{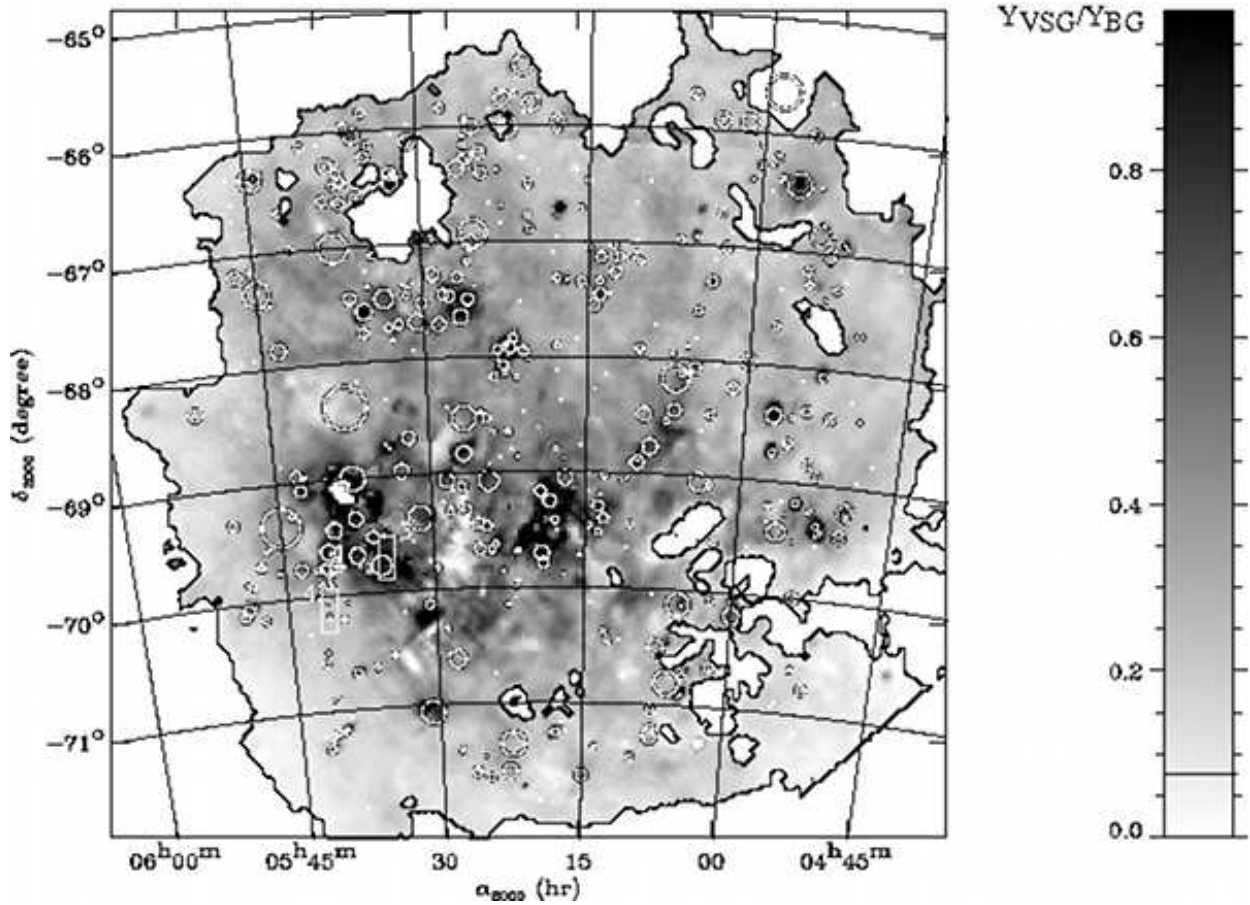}
\caption{Map of the VSG relative abundance with respect to BGs. The
overlaid symbols show the HII regions \citep[][catalog]{Davies76}.
Regions 3 and 4 indicated by rectangles are described in
Table \ref{table_regions}. The horizontal line in the color bar shows the value of the standard MW relative abundance.\label{fig_vsg_bg} }
\end{center}
\end{figure*}

\begin{figure*}
\begin{center}
\includegraphics[width=7.5cm]{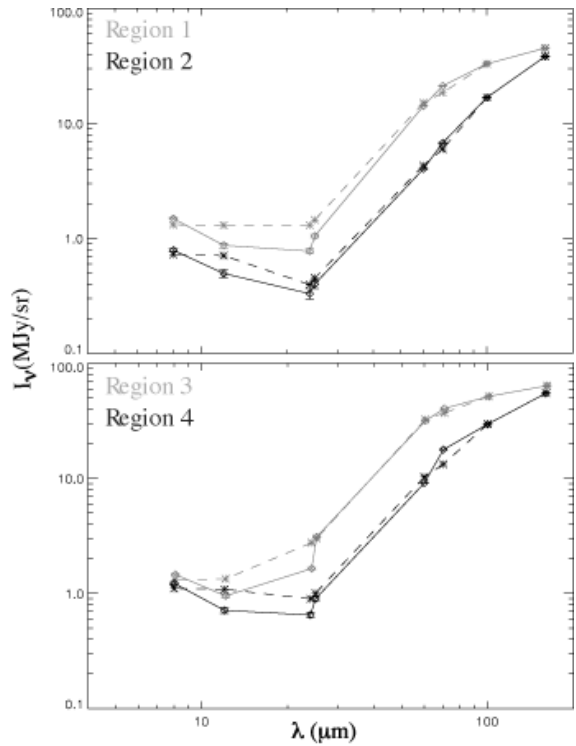}
\caption{Median brightnesses between 5.8 and 160 $\mic$ deduced from
the data (diamond linked by continuous line) and brightnesses deduced
from the model (stars linked by dashed line). The top panel
corresponds to SEDs in regions 1 and 2 (see Figure \ref{fig_pah_bg})
with a low and high PAH relative abundance in black and gray
respectively. The bottom panel corresponds to values in regions 3 and
4 (see Figure \ref{fig_vsg_bg}) with low and high VSG relative
abundance in black and gray respectively. \label{fig_spec_regions} }
\end{center}
\end{figure*}

Figure \ref{fig_vsg_bg} shows the VSG relative abundance map. The VSG
distribution across the LMC is different from that of the
PAHs. Regions with enhanced VSGs are located around \DOR and in a patch
near the center of the LMC. As for he PAHs, the increase of the
VSG abundance near the center of the bar cannot come from a reddened
ISRF, since this would produce a decrease instead of an increase of
their abundance. As opposed to the PAH abundance, the VSG relative
abundance follows the distribution of the HII regions quite well
(symbols in Figure \ref{fig_vsg_bg}). Note that a similar artifact as
in the PAH abundance is evidenced in the north-south region between
\DOR and the bar (see Section \ref{sec_pah} for details).

The bottom panel of Figure \ref{fig_spec_regions} shows the median
spectra for two regions (regions 3 and 4) with low and high VSG
relative abundance ($\rm Y_{VSG}/Y_{BG}=0.24$ and 1.18) respectively
and their corresponding models.  The main difference between the two
spectra in the FIR domain (from 60 to 160 $\mic$) is the
flatness of the spectrum in region 3 with high VSG relative
abundance. The $\rm \alpha_{VSG}$ value is equal to 1 for the model
spectra, but the derived VSG relative abundance is increased by a
factor of 2 from region 4 to 3.  As discussed in Section
\ref{sec:desert}, we only allow two values for the VSG size
distribution slope, $\rm \alpha_{VSG}$=1 or the Galactic value, 2.6. For
most pixels, the value deduced from the fit is 1, but some regions
require the Galactic value. This is the case of high $\XISRF$ regions
such as \DOR and some individual HII regions.  In these warm
regions, the 70 $\mic$ emission is actually dominated by warm BG
emission and little additional emission from VSGs is required.  In
those cases, the model with $\rm \alpha_{VSG}=2.6$ sometimes overestimates
the 70 $\mic$ emission, which could indicate the presence of a steeper
VSG size distribution than in the MW.  In summary, results of the fits
give $\rm \alpha_{VSG}$ = 1 in the overall LMC except for HII regions
where the Galactic value (or higher value) is needed.

\subsubsection{ISRF mixture results}

The $\rm \alpha$ parameter (radiation field probability distribution)
used in Equation \ref{eq_dale} ranges from 1 to 2.5. \cite{Dale01}
describe the two extreme cases : $\rm \alpha$=2.5 corresponds to diffuse
regions, whereas $\rm \alpha$=1 represents photodissociation
regions near young stars. In the LMC, regions with $\rm \alpha$=2.5 are
more common. The decrease of the parameter $\rm \alpha$ across the LMC
follows the increase of the BG equilibrium temperature \citep[See
Figure 7 of][]{Bernard08}.

\begin{figure*}
\begin{center}
\includegraphics[width=16.5cm]{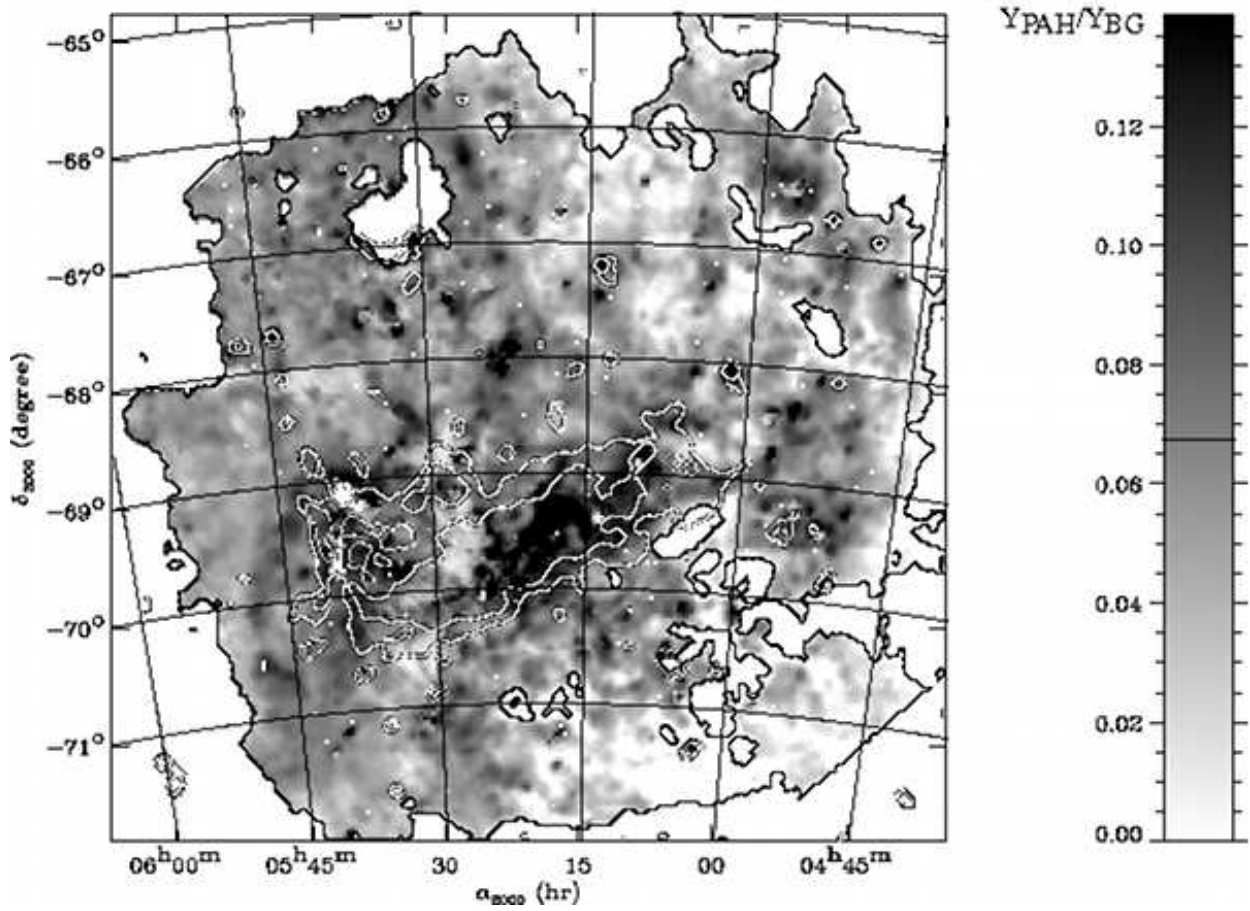}
\caption{Map of the PAH relative abundance with respect to BGs, for the
case of an ISRF mixture along the line of sight. Contours correspond
to the Two Microns All Sky Survey (2MASS) J-band map, convolved to the
4$^{\prime}$ resolution. The horizontal line in the color bar shows the value of the standard MW relative abundance.\label{fig_pah_dale}}
\end{center}
\end{figure*}

\begin{figure*}
\begin{center}
\includegraphics[width=16.5cm]{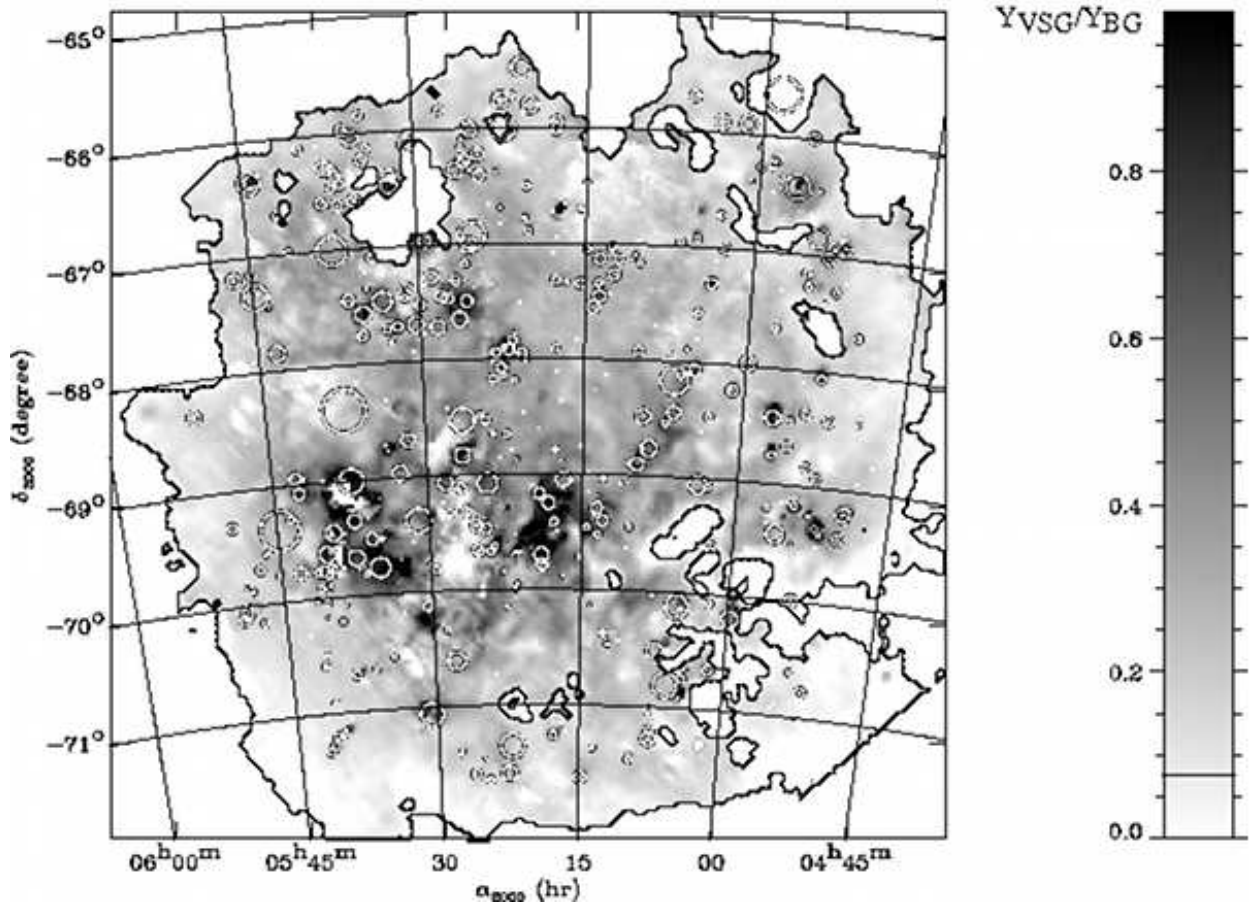}
\caption{Map of the VSG relative abundance with respect to BGs, for the
case of an ISRF mixture along the line of sight. The overlaid symbols
show the HII regions (Davies et al, 1976 catalog). The horizontal line in the color bar shows the value of the standard MW relative abundance.\label{fig_vsg_dale}}
\end{center}
\end{figure*}

The PAH and VSG relative abundances derived for the case where we let
the ISRF strength mixture to vary are given in Figures
\ref{fig_pah_dale} and \ref{fig_vsg_dale}, respectively. The scale in
the two maps are the same as in Figures \ref{fig_pah_bg} and
\ref{fig_vsg_bg}. Figure\,\ref{fig_pah_dale} still shows the increase
of the PAH abundance in the old-population stellar bar, although the
amplitude of the variation is slightly less than that in Figure
\ref{fig_pah_bg}. The VSG relative abundance map is quite similar to
the one shown in Figure \ref{fig_vsg_bg} : the VSG relative abundance
still follows roughly the distribution of the HII regions and the main region
of enhanced VSG abundance is still located around \DOR and near the
center of the LMC. Results do not appear to change significantly when
we consider the effect of a realistic ISRF strength mixture along the
line of sight.

\section{Discussion}
\label{sec:discussion}

PAHs and VSGs could have a different origin or could be subjected
to different processing in the ISM. Several studies of
the MIR aromatic bands and continuum emission around 15 $\mic$
showed that PAH and VSG emission are linked to star formation
activity. \cite{Forster04} describe the PAH component as
representative of the quiescent environments while the VSG
component traces the active star formation sites. They attribute the
disappearance of PAH bands in HII regions to the destruction of PAHs
in regions with high intensity of the radiation field \citep[in agreement
with the decrease of the IRAS 12/100 ratio in regions of intense UV
radiation, from][]{Boulanger88}. \cite{Povich07} considered two
possible mechanisms for PAH destruction in HII regions : stellar UV
photons and X rays. They concluded that only UV photons could be
responsible for the absence of PAHs in these regions.
\cite{Contursi00} examined the N4 region in the LMC using
ISOCAM data and found that the maximum emission is located at the
interface between the HII region and the molecular cloud, as was also
found from observations of M17 in our Galaxy by \cite{Cesarsky96}.
\cite{Helou04} studied the spiral galaxy NGC 300 using Spitzer data
and found that emission at 24 $\mic$ traces the star-forming regions
whereas the emission at 8 $\mic$ highlights the edges of these
regions. However, interpreting those observations in terms of VSG
abundance remains difficult, since the excitation conditions are
generally poorly constrained.

At least a fraction of the PAH population is thought to be formed and
injected into the ISM by the stellar winds following mass loss from
old carbon-rich stars. \cite{Loup99} conducted a survey of 0.5 square
deg$^2$ in the bar of the LMC using the ISOCAM and DENIS
instruments. Their objective was to build a complete sample of
thermally pulsing asymptotic giant branch (TP-AGB) stars. During this
phase stars lose mass and carbon-rich AGB could contribute
significantly to the formation of PAHs. They found approximately 300
TP-AGB stars, of which 9$\%$ have high mass-loss rates. Some studies
of post-AGB stars in transition between the AGB and planetary nebulae
phase exhibit signatures of dust attributed to PAHs
\citep[e.g.,][]{Raman08}. \cite{Justtanont96} analyzed MIR
spectroscopy in three post-AGB stars, known to exhibit an emission feature
at 21 $\mic$. In addition to a weak emission at 11.3 $\mic$, they found
several features which match the emission bands in the spectrum of
chrysene (C$_8$H$_{12}$), which is one of the simplest PAH
molecules. The 21 $\mic$ emission feature has never been observed in
carbon rich AGB stars, nor planetary nebulae. The PAH molecule responsible for this feature could
be formed in the dust shells around some post-AGB stars and lead to
increase the abundance of this PAH species \citep{Justtanont96}.

In this study, we showed that the PAH relative abundance in the LMC
is maximum in an extended region corresponding to the old-population
stellar bar, a region with no strong current stellar formation
activity.  To our knowledge, this is the first time such an effect is
evidenced at Galactic scale. We interpret this finding as an
indication that PAHs may have formed efficiently in this region in the
past from carbon-rich stars winds and have remained overabundant in
the surrounding ISM. Our result also shows the presence of a region
with a significantly increased VSG abundance in the same region,
although with a lesser spatial extent. This is likely to reveal that VSGs
also formed efficiently from the same process. The difference in the
spatial extent of the regions could either reflect specific formation
processes or differences between the evolution of the two species in
the ISM. In the later hypothesis, our result would suggest that PAHs
are in fact more resistant than VSGs to destruction processes in the
diffuse ISM.

\begin{figure*}
\begin{center}
\includegraphics[width=16.5cm]{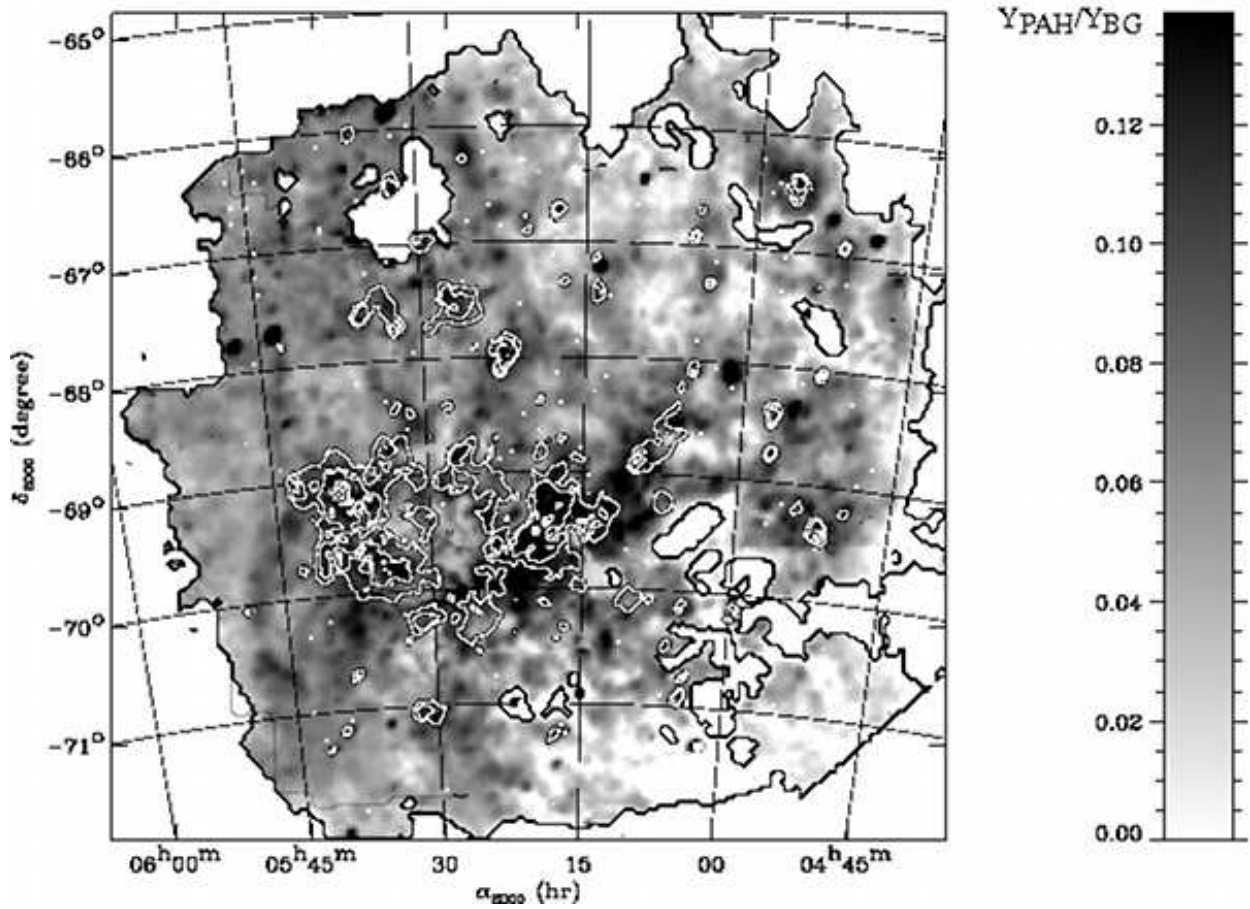}
\caption{Map of the PAH relative abundance with respect to BGs, overlaid with contours at 0.6 and 0.9 of the VSG relative abundance. \label{fig_pah_ct_vsg}}
\end{center}
\end{figure*}

A careful examination of Figure \ref{fig_pah_ct_vsg} which compares the spatial
distribution of PAHs and VSGs indicates that there are other regions
where the PAH and VSG abundances are spatially correlated. Most
noticeably, the region surrounding \DOR and its extension to the
southwest both have enhanced PAH and VSG abundances. However,
significant differences are also apparent. For instance, the region of
enhanced PAH abundance to the west of the main molecular ridge, South
of \DOR, as well as the arc-like molecular ridge to the east have no
counterpart in the VSG abundance map. Similarly, at the northwest end
of the stellar bar, the PAH and VSG abundance maps appear
anticorrelated.

In the solar neighborhood, such independent variations of the
distribution of PAHs and VSGs have been observed. For instance,
\cite{Bernard93} showed the existence of increased PAH abundance
regions at the periphery of some translucent clouds. This could result
from the destruction of larger species such as VSGs into PAH
molecules, induced by the action of external low-velocity shocks at
the surface of the clouds. \cite{Mamd02} evidenced large variations
of the PAH emission in the Ursa Major cirrus.
They proposed that the observed color variations are due
to actual PAH abundance variations due to grain shattering in
energetic grain-grain collisions induced by turbulence or grain
coagulation caused by small turbulent motions. Therefore regions with
large turbulent motion could generate PAHs through shattering of
larger dust particles, such as VSGs. Indeed, the destruction of
graphitic dust is expected to lead to the formation of large two
dimensional PAH-like molecules. \cite{Berne07} studied a
photodissociation region using the Spitzer IRS data and found a
disappearance of VSG emission while the PAH emission increases. Like
\cite{Cesarsky00} they suggested the possibility that VSGs could
evolve into PAHs under UV photon driven processes. Under this
hypothesis, VSGs can be viewed as PAH clusters.  Although at the
distance of the LMC, such small-scale structure would not be resolved,
it is likely that the differences we evidence between the spatial
distribution of PAHs and VSGs are of a similar nature.  It is worth
noticing that, in areas with no VSG enhancement such as the southern
tip of the arc-shaped molecular ridge, the PAH abundance enhancement
appears to trace the surface of the molecular clouds, a similar
situation as that observed in nearby halo clouds, although with much
larger physical scale, given the distance of the LMC.

Similar variations were observed in the SMC by \cite{Bolatto06} who
found large variations of the $\rm I_{8}/I_{24}$ ratio across the
southwest region of the bar. They attributed these to spatial
variations of the PAH abundance, with the PAH abundance decreasing
with increasing 24 $\mic$ brightness. They suggested that the PAH
abundance variations are essentially driven by the local ISRF and PAH
photodestruction.  In the LMC, \cite{Bernard08} showed that a large
region around of \DOR has excess 70 $\mic$ emission, which they
interpreted through the flattening of the VSG size distribution,
potentially caused by the erosion of larger BGs by shocks propagating
from the \DOR region.  In our study, we find that this same region
also has an increased VSG abundance. Our result shows that this
region, in addition to showing an increase of the largest VSGs also
shows an increase of the overall VSG abundance.  \cite{Bernard08}
indicated that the increase of the largest VSG abundance would
correspond to $\simeq\,13\%$ of the BG mass, which converts into a VSG
abundance increase of 77\% for a standard MW dust mixture. The excess
we measure in this region corresponds to an increased abundance by a
factor of about 2.4, and is therefore larger than that imputable to VSG
size distribution change only. Similarly,
\cite{Lisenfeld02} analyzed the emission from NGC 1569, a dwarf
galaxy with a metallicity between those of the LMC and the SMC. Using
the \cite{Desert90} model, they evidenced an enhancement of the VSG
abundance relative to BGs, with respect to the solar neighborhood, by
a factor of 2-7, depending on the model parameters they used. They
explain this result by the BG destruction due to shocks in the
turbulent ISM of this galaxy.

We therefore consider it likely that the differences evidenced here between
the spatial distribution of the PAH and VSG abundances relative to BGs
are tracing both the production of PAHs from the destruction of VSGs at
the surface of molecular clouds, and the production of VSGs from
the destruction of larger grains in shocks.

\begin{figure*}
\begin{center}
\includegraphics[width=7.5cm]{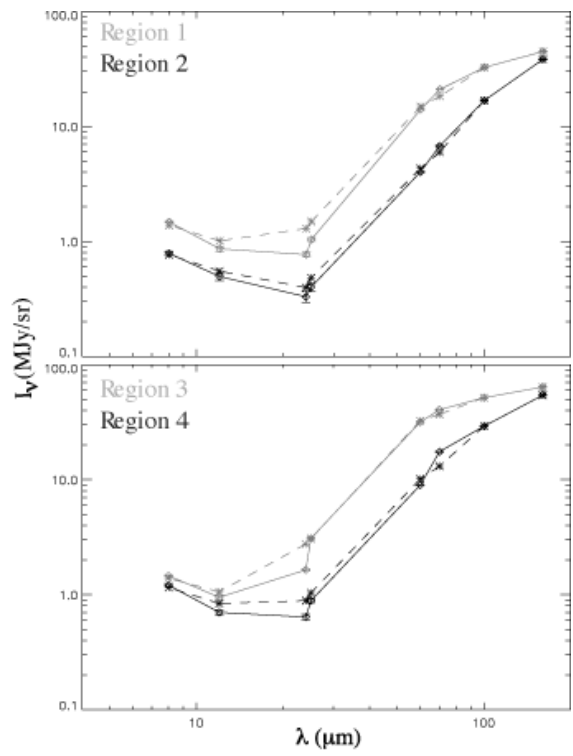}
\caption{Same figure as Figure \ref{fig_spec_regions}, using a model
with 100$\%$ ionized PAHs. \label{fig_spec_regions_pahion} }
\end{center}
\end{figure*}

We saw in Figure \ref{fig_spec_regions} that the model with neutral
PAHs was not able to reproduce the IRIS data at 12 $\mic$. We suspect
that this is due to the ionization state of the PAHs or the nature of
the molecules, such as dehydrogenation. It is difficult to constrain
the nature of the PAHs in the LMC from the data presented here
only. However, varying the PAH ionization fraction in the model, we
find that the model requires 100$\%$ of the PAHs in the ionized state
to explain the observations at 12 $\mic$ (see Figure
\ref{fig_spec_regions_pahion}). Even if this model does not reproduce
the data at 12 $\mic$ perfectly, the difference between the model and
the data is smaller than with our initial hypothesis. \cite{Flagey06}
obtained a neutral over ionized PAHs ratio close to 50 $\%$ in the
MW (GLIMPSE field). Therefore, PAHs in the LMC could be more
ionized than in the Galactic plane. However, the exact nature of
PAHs in the LMC could be different than in our Galaxy. A study of the
nature of the PAH molecules is essential to determine the origin of
the observed difference at 12 $\mic$ between our Galaxy and the
LMC. Such study may become possible using the data from the Spitzer
Sage-Spec legacy program.

\section{Conclusion}
\label{sec:cl}

We used the data of the SAGE Spitzer legacy survey, combined to the
IRIS data to measure the variations in the colors of infrared emission from dust in the LMC. Comparisons with results
from the literature showed a higher $\rm I_8/I_{24}$ ratio in the LMC
than in other low-metallicity environments, which could result from a
relatively higher PAH abundance in the LMC with respect to its
metallicity class. Morever, the $\rm I_{3.6}/I_{8}$ and $\rm
I_{4.5}/I_{8}$ ratios are smaller in the LMC than in the MW. This could result from different properties of the near-infrared
continuum or from different properties of the PAHs. In
particular, PAHs could be smaller in the LMC, compared to our Galaxy,
and more ionized. The SED modeling of regions with enhanced PAH and
VSG relative abundances in the LMC show a similar result : the model
used requires a large fraction (potentially 100$\%$) of ionized PAHs
in order to minimize the difference between the model and the data at
12 $\mic$.

In order to understand the processes of grain formation and
destruction at the scale of the LMC, we constructed for the first time
maps of the PAH and VSG relative abundances with respect to BGs of the
overall galaxy, taking into account the distribution of the radiation
field intensity, as inferred from the equilibrium temperature of the
BGs. We evidenced an increase of the PAH relative abundance over an
extended region which matches the extent of the old-population stellar
bar. PAHs are also overabundant at the periphery of some molecular
clouds. The VSG abundance is increased over a region around \DOR
and in a small region near the center of the bar. This clear
distinction between the PAH and VSG spatial distribution could be due
to different origins or different processings in the ISM for these two
dust components. In agreement with several studies, the VSG abundance
distribution appears to follow the star formation activity while that
of PAHs could follow more quiescent environments.

We interpret the larger PAH abundance in the bar as a sign of the past
formation of these species by C-rich mass losing stars.  VSGs could
also be formed from the same process but the difference in the spatial
distributions remains unclear : the two dust components could originate
from different stellar populations with different distribution, or VSGs
and PAHs could have been evolved differently in the diffuse ISM. If
destruction is at work, this may indicate that VSGs are actually
destroyed faster than PAHs in the diffuse ISM.  We interpret the rest
of the abundance variations as signs of the evolution of dust species
of different sizes in the ISM.  The region around \DOR shows
increased VSG abundance, which could result from the destruction of
larger BGs in shocks.  Similarly, regions close to the surface of
molecular clouds with an increase PAH relative abundance could result
from the destruction of VSGs into PAHs by shocks impacting the cloud's
surfaces.


\begin{table*}
\caption{Infrared color ratios of data. \label{tab_ir_color}}
\begin{center}
\begin{tabular}{ c c c c c c }
\hline
\hline
IR ratios & LMC  & LMC$\rm ^{a}$ & SMC$\rm ^{b}$ & GC$\rm ^{d}$  & 34 galaxies$\rm ^{d}$ \\
          &      &       &    & GFLS/ GLIMPSE$\rm ^{e}$ & \\
\hline
$\rm I_{100}/I_{160^{f}}$ & $(6.6 \pm 0.5) \,10^{-1}$ & $(7.0 \pm 1.9) \,10^{-1}$ & 1.0 $\pm$ 0.2 &  - & - \\
$\rm I_{70}/I_{160}$  & $(4.5 \pm 0.8) \,10^{-1}$ & $(4.5 \pm 0.9) \,10^{-1}$ & - & - & - \\
$\rm I_{60}/I_{160^{f}} $ & $(3.1 \pm 0.7) \,10^{-1}$ & $(3.4 \pm 0.7) \,10^{-1}$ & 0.5 $\pm$ 0.1 & - & - \\
$\rm I_{25}/I_{160^{f}}$ & $(2.4 \pm 1.0) \,10^{-2}$ & $(1.9 \pm 0.6) \,10^{-2}$ & $(2.0 \pm 1.0) \,10^{-2}$ &  - & - \\
$\rm I_{24}/I_{160}$ & $(1.7 \pm 0.3) \,10^{-2}$ & $(2.5 \pm 0.5) \,10^{-2}$ & - &  - & - \\
$\rm I_{12}/I_{160^{f}}$ & $(1.5 \pm 0.4) \,10^{-2}$ & $(7.4 \pm 3.8) \,10^{-3}$ & $(1.1 \pm 0.6) \,10^{-2}$ & - & - \\
$\rm I_{8}/I_{160}$ & $(2.5  \pm 0.2) \,10^{-2}$ & $(1.6 \pm 0.4) \,10^{-2}$ & - &  - & - \\
$\rm I_{5.8}/I_{160}$ & $(9.2 \pm 7.1) \,10^{-3}$ & $(2.7 \pm 0.9) \,10^{-3}$ & - &  - & - \\
\hline
$\rm I_{8}/I_{24}$ & $ 1.6 \pm 0.3$ & $(6.3 \pm 1.4) \,10^{-1}$ & 0.36$\rm ^{g}$ &  - & $(8.0 \pm 4.0) \,10^{-2}$$\rm ^h$ \\
$\rm I_{8}/I_{12}$  & 1.7 $\pm$ 0.2 & 2.1 $\pm$ 1.1 & - &  - & - \\
$\rm I_{5.8}/I_{8}$ & $(3.3 \pm 2.1) \,10^{-1}$ & $(1.7 \pm 0.6) \,10^{-1}$ & - & $(3.7 \pm 0.3) 10^{-1}$$\rm ^{c}$  & - \\
   &      &    &  & $(2.6 - 3.7) \,10^{-1}$$\rm ^{e}$ & \\
$\rm I_{4.5}/I_{5.8}$ & $(5.4 \pm 7.7) \,10^{-1}$ & $(4.3 \pm 28.5) \,10^{-2}$ & - & - & -\\
$\rm I_{3.6}/I_{4.5}$ & $ (1.6 \pm 0.2) \,10^{-1}$ & 1.1 $\pm$ 7.6 &- &  - \\
\hline
$\rm I_{4.5}/I_{8}$ & $ (1.8 \pm 3.7) \,10^{-1}$ & $(7.4 \pm 5.6) \,10^{-3}$ &  - & $(3.7 - 6.5) \,10^{-2}$$\rm ^{e}$ & $>$0.1 if $\rm I_{8}/I_{24} < 0.1$$\rm ^h$\\
$\rm I_{3.6}/I_{8}$ & $ (2.9 \pm 6.3) \,10^{-3}$ & $(7.9 \pm 7.6) \,10^{-3}$&  - & $(5.9 - 9.4) \,10^{-3}$$\rm ^{e}$ & -\\
\hline
\end{tabular}
\end{center}
$\rm ^{a}$ \cite{Bernard08} (overall LMC) \\
$\rm ^{b}$ \cite{Bot04} (overall SMC)\\
$\rm ^{c}$ Galactic Center from \cite{Arendt08} \\
$\rm ^{d}$ \cite{Engelbracht05} \\
$\rm ^{e}$ Galactic First Look Survey / Galactic Legacy Infrared Mid-Plane Survey Extraordinaire from \cite{Flagey06} \\
$\rm ^{f}$ 170 $\mic$ instead of 160 $\mic $ for the \cite{Bot04} ratios \\
$\rm ^{g}$ \cite{Bolatto06} (Southwest Bar region of the SMC) \\
$\rm ^{h}$ In subsolar metallicity environments \citep[with a solar metallicity reference 12+log(O/H)=8.7,][]{Prieto01}

\end{table*}

\begin{table*}[!ht]
\caption{SEDs and Results of the SED Fits for the Four Selected Regions :
1, High $\rm \frac{Y_{PAH}}{Y_{BG}}$; 2, Low $\rm
\frac{Y_{PAH}}{Y_{BG}}$; 3, High $\rm \frac{Y_{VSG}}{Y_{BG}}$; 4, Low
$\rm \frac{Y_{VSG}}{Y_{BG}}$. \label{table_regions}}
\begin{center}
\begin{tabular}{lcccc}
\hline
\hline
Region & 1   & 2 & 3 & 4 \\
\hline
$\rm \alpha_{2000}$ & 05:23:00 & 05:43:00 & 05:35:00 & 05:42:00 \\
$\rm \delta_{2000}$&  -69:30:00 & -71:10:00 & -69:47:00 & -70:15:00 \\
Box size ($\rm \degr$) & 0.4 $\times$ 0.4 & 0.4 $\times$ 0.4 & 0.4 $\times$ 0.4 & 0.4 $\times$ 0.4 \\
I$_{160}$ (MJy sr$^{-1}$)  & 46.72 $\pm$ 0.49  & 40.30 $\pm$ 0.47 & 66.05 $\pm$ 0.51 & 56.78 $\pm$ 0.48 \\
I$_{100}$ (MJy sr$^{-1}$)  & 34.06 $\pm$ 0.23 & 17.68 $\pm$ 0.22 & 53.34 $\pm$ 0.24 &30.51 $\pm$ 0.22 \\
I$_{70}$ (MJy sr$^{-1}$)  & 21.91 $\pm$ 0.10 & 7.12 $\pm$ 0.09 & 41.90 $\pm$ 0.10 & 18.32 $\pm$ 0.09 \\
I$_{60}$ (MJy sr$^{-1}$)  & 14.67 $\pm$ 0.05 & 4.23 $\pm$ 0.05 & 32.70 $\pm$ 0.05 & 9.38 $\pm$ 0.05 \\
I$_{25}$ (MJy sr$^{-1}$)  & 1.08 $\pm$ 0.04 & (4.23 $\pm$ 0.36) 10$^{-1}$ & 3.20 $\pm$ 0.04 & (9.28 $\pm$ 0.36) 10$^{-1}$ \\
I$_{24}$ (MJy sr$^{-1}$)  & (8.02 $\pm$ 0.37) 10$^{-1}$ & (3.48 $\pm$ 0.37) 10$^{-1}$ & 1.70 $\pm$ 0.04 & (6.69 $\pm$ 0.37) 10$^{-1}$ \\
I$_{12}$ (MJy sr$^{-1}$)  & (9.01 $\pm$ 0.44) $10^{-1}$ & (5.22 $\pm$ 0.42) 10$^{-1}$ & (9.81 $\pm$ 0.43) 10$^{-1}$ & (7.34 $\pm$ 0.42) 10$^{-1}$ \\
I$_{8}$ (MJy sr$^{-1}$)  & 1.54 $\pm$ 0.02 & (8.31 $\pm$ 0.22) 10$^{-1}$ &1.51 $\pm$ 0.02 & 1.26 $\pm$ 0.02 \\
$\rm Y_{PAH}/Y_{BG}$ & 1.41 10$^{-1}$ & 7.27 10$^{-2}$ & 1.37 10$^{-1}$ & 8.31 10$^{-2}$ \\
$\rm Y_{VSG}/Y_{BG}$ & 4.87 10$^{-1}$ & 1.22 10$^{-1}$ & 1.09 & 2.28 10$^{-1}$ \\
X$\rm _{ISRF}$ & 0.85 & 0.49 & 0.28 & 0.59 \\
\hline
\end{tabular}
\end{center}
\end{table*}

\end{document}